**Title**: Digital citizen science for ethical surveillance of physical activity among youth: mobile ecological prospective assessments vs. retrospective recall

**Authors**: Sheriff Tolulope Ibrahim[1], Jamin Patel[1, 2], Tarun Reddy Katapally[1, 2, 3]

**Affiliations**:

[1]DEPtH Lab, School of Health Studies, Faculty of Health Sciences, Western University, London, Ontario, N6A 3K7, Canada; tarun.katapally@uwo.ca.

[2] Department of Epidemiology and Biostatistics, Schulich School of Medicine and Dentistry, Western University, London, Ontario, N6A 3K7, Canada; tarun.katapally@uwo.ca.

[3]Lawson Health Research Institute, 750 Base Line Road East, Suite 300, London, Ontario, Canada N6C 2R5

**Corresponding author**:
Tarun Reddy Katapally
School of Health Studies
Faculty of Health Sciences
Western University
1151 Richmond St
London, ON N6A 3K7.
Phone: +1(519) 661-4249
Email: tarun.katapally@uwo.ca

**Abstract**

Physical inactivity is the fourth leading risk factor of mortality globally. Hence, understanding the moderate-to-vigorous physical activity (MVPA) patterns of youth is essential to manage and mitigate non-communicable diseases. As digital citizen science approaches utilizing citizen-owned smartphones to ethically obtain MVPA big data can transform physical activity (PA) surveillance, this study aims to understand the frequency of MVPA reported by youth using smartphone-deployed retrospective validated surveys (adapted from the International Physical Activity Questionnaire, the Simple Physical Activity Questionnaire, and the Global Physical Activity Questionnaire), compared to prospective time-triggered mobile ecological prospective assessments (mEPAs). Using a digital citizen science methodology, this study recruited youth citizen scientists ($N = 808$) in 2018 (August 31 to December 31) in Saskatchewan, Canada. A total of 68 participants (aged 13 to 21) provided complete data on prospective mEPAs and retrospective surveys. Statistical analyses included a Wilcoxon signed-rank test to compare retrospective and prospective MVPA frequency and Poisson regression, adjusting for age, gender, parental education, and ethnicity to explore associations between various contextual factors and frequency of MVPA. A significant difference was found in reporting the frequency of MVPA retrospectively vs. prospectively via mEPAs ($p < 0.000$). Family and friends support for MVPA, drug free friends, part time job, and distance were associated with prospective MVPA frequency; however, only school and strength training displayed a significant association with retrospective MVPA frequency. With access to ubiquitous digital devices growing worldwide, and youth having particularly high digital literacy, digital citizen science for the ethical surveillance of PA using mEPAs presents a promising approach for effective PA surveillance and designing effective PA recommendations for youth.

## Author Summary

Physical inactivity is a major global health concern and a leading risk factor for death worldwide. To better understand and manage non-communicable diseases, it's important to monitor how often young people engage in moderate-to-vigorous physical activity (MVPA). This study is part of the Smart platform, a study that combines citizen science, community based participatory research and systems science to engage participants. This study compared two methods of MVPA surveillance (i.e., retrospective surveys versus real-time mobile ecological prospective assessments - mEPAs). In 2018, 68 youth aged 13 to 21 in Saskatchewan, Canada, used their smartphones to report their MVPA and other health related behaviour over an eight-day study period, including weekdays and weekends. This data was then used to determine the frequency of MVPA. The study found significant differences in the frequency of MVPA reported retrospectively versus mEPAs. Additionally, variations were observed in how contextual and demographic factors were associated with physical activity in both types of assessments. This suggests the need to adapt physical activity monitoring practices to leverage digital tools and engage participants through their own devices for more comprehensive surveillance.

## Introduction

Evidence is clear that physical activity (PA) is crucial in the prevention and management of non-communicable diseases (NCD) among youth [1], yet over 80% of youth do not meet the recommended PA guidelines [2]. PA has substantial health benefits, with evidence indicating the presence of a dose-response relationship between PA and health – the more PA that youth engage in, the greater their health benefits [3]. For instance, studies have shown that youth participation in PA is associated with improved cardiovascular fitness [4], decreased risk of type 2 diabetes, improved bone health [5], reduced body fat, and stronger muscles [3,6,7]. Additionally, engagement in consistent PA has been found to improve sleep quality, reduce symptoms of depression and anxiety, and minimize substance use [8]. Numerous studies have also found that PA is associated with improved self-reported health status [9], and lower odds of obesity, with moderate intensity PA demonstrating a particularly strong association compared to light intensity PA [3].

Moderate to vigorous physical activity (MVPA) is a complex and multifaceted behaviour that requires accurate measurement across various dimensions, including intensity, frequency, duration, and mode [10]. Accurate measurement of MVPA is essential for monitoring population adherence to PA recommendations, and for developing policies and programs that promote PA. For instance, the Centre for Disease Control and Prevention recommends that children and adolescents should engage in 60 minutes of MVPA per day, with three of those days being vigorous-intensity PA [11]. Measuring MVPA frequency enables direct comparisons with established guidelines, making it easier to assess adherence to recommended levels of PA [12]. However, surveillance of MVPA poses a substantial challenge due to inherent limitations in existing measurement methodologies, which are susceptible to measurement bias, primarily stemming from recall errors [13,14] and social desirability [15]. Recall errors introduce inaccuracies in reported MVPA data [16], while the influence of social desirability [17], particularly apparent in subjective recall measures, can further distort the reliability of MVPA data [18]. On the other hand, objective measures, while providing continuous, reliable, and passive assessments of MVPA [19], have their own limitations [20]. Objective measures lack the capacity to capture contextual and environmental factors that act as both facilitators and barriers to MVPA [21]. This limitation hinders a comprehensive understanding of population MVPA measurement. Moreover, objective measures are resources intensive and logistically more difficult to deploy [22,23]. Repeated prospective measures can address these deficiencies of existing methodologies, particularly when deployed via citizen-owned digital devices to minimize recall bias and enhance collection of environmental factors that influence MVPA [24]. Hence, the rationale for carrying out this study to understand the frequency of MVPA reported retrospectively through a modified survey and prospectively through prospective assessments in the same cohort of participants.

As the Social Cognitive Theory posits that personal, social (i.e., friends, adult support etc.), and environmental factors shape human behaviour, objective measures may not be able to provide a comprehensive understanding of MVPA among youth, potentially overlooking factors that can act as facilitators and barriers to MVPA [25]. For instance, during the youth stage of development, there is a notable increase in time spent with peers compared to parents, leading to a significant impact on the formation of norms and behaviours [26,27]. Social factors such as peers and parent involvement in PA have been closely linked to increased PA among children and adolescents [27–31]. Moreover, the importance of capturing contextual factors as it affects youth PA has been established [32]. For example, engagement in muscular strength training has been reported to be associated with physical, psychosocial, emotional, and behavioural aspects that contribute to PA in youth [6]. Additionally, participation in PA has been found to be associated with less illicit drug usage among youth [33]. In this study, social factor associated with MVPA was captured by peer support for PA [34], family support for PA [35] and friends committed to being drug-free. In another study, distance between home and school have been found to be an important predictor for active travel which in turn contributes to MVPA accumulation [36].

Digital citizen science can transform the ethical surveillance of MVPA among youth [37]. The rapid growth and near-universal access to ubiquitous devices such as smartphones in the Western world [38] has provided more opportunities to engage youth as active participants (i.e., citizen scientists) in data collection [39]. For instance, mobile ecological momentary assessments deployed through youths' own smartphones can assess accurate MVPA behaviour in real-time and real-world situations [40], while capturing different participation contexts (mode, settings and context) [41] and dimensions (frequency, duration, intensity and type) [42]. However, the implementation of conventional ecological momentary assessments for real-time data collection of MVPA may face challenges with compliance due to assessment schedules [43]. In this study, the use of digital citizen science facilitated valuable feedback from participants who informed our research team that promptly responding to momentary assessments were challenging due to their busy daytime schedules and difficulties in recording MVPA as it occurs in real-time. To address these concerns, we modified the momentary assessments to encompass a longer reporting interval (i.e., end of day), following Shiffman et al [44]. Consequently, throughout this study, our adapted assessments will be referred to as mobile Ecological Prospective Assessments (mEPAs), to distinguish our approach from traditional ecological momentary assessments.

While several studies have compared retrospective and momentary assessment of human behaviour including sedentary behaviour [45], PA [46], and substance use [47] in adult populations, there is a dearth of evidence examining how youth engage with smartphones to report their MVPA using both digitally deployed retrospective surveys as well as mEPAs. Following the shortcomings of retrospective recall measure (i.e., recall, self-reporting, and social desirability biases), we hypothesize that these shortcomings could be reduced, if not completely avoided, by adopting mEPAs. The accessibility model supports the use of prospective assessments to avoid recall bias [48]. This model posits that self-reports with longer recall periods will be based on beliefs and implicit theories in comparison to self-reports with shorter recall periods (i.e., self-reports within the same day) [49]. This study aims to address this gap by engaging youth as citizen scientists [50] to compare MVPA frequency reported via digitally-deployed retrospective surveys and prospective mEPAs within the same cohort of participants. The study also aims to identify the sociodemographic, behavioural, and contextual factors associated with the frequency of MVPA reported by youth retrospectively vs. prospectively through mEPAs.

## Methods

### Study Design

The study included both cross-sectional and longitudinal measures [24,51] to engage with participants in Regina, an urban centre located in the Canadian prairie province of Saskatchewan. This study examined MVPA behaviours and associated factors among youth who participated in the Smart Platform [39]. The Smart Platform combines citizen science, community-based participatory research, and systems science to digitally explore behavioural phenomena, engage in knowledge translation, and deploy real-time interventions [39,52] in a given population. At the core of the Smart Platform are the citizen scientists who engage with the platform through their personally owned smartphones for population health research. Ethics approval for the Smart Platform was granted by the Research Ethics Board of the University of Regina and Saskatchewan (REB #2017-029).

The Smart Platform (**SI Figure 1**) links a frontend smartphone application (app) deployed on citizen scientists' ubiquitous devices with a backend dashboard accessed by academic scientists. Essentially, the Smart Platform consists of two key human-computer interaction interfaces: citizen interface (smartphone app) and scientist interface (dashboard where anonymized citizen-generated big data are visualized in real-time). More importantly, the Smart Platform enables remote interaction in real-time, where the citizen scientists and/or academic scientists can initiate anonymized communication.

**S1 Figure 1**: The Smart Platform: citizen interface + scientist interface.

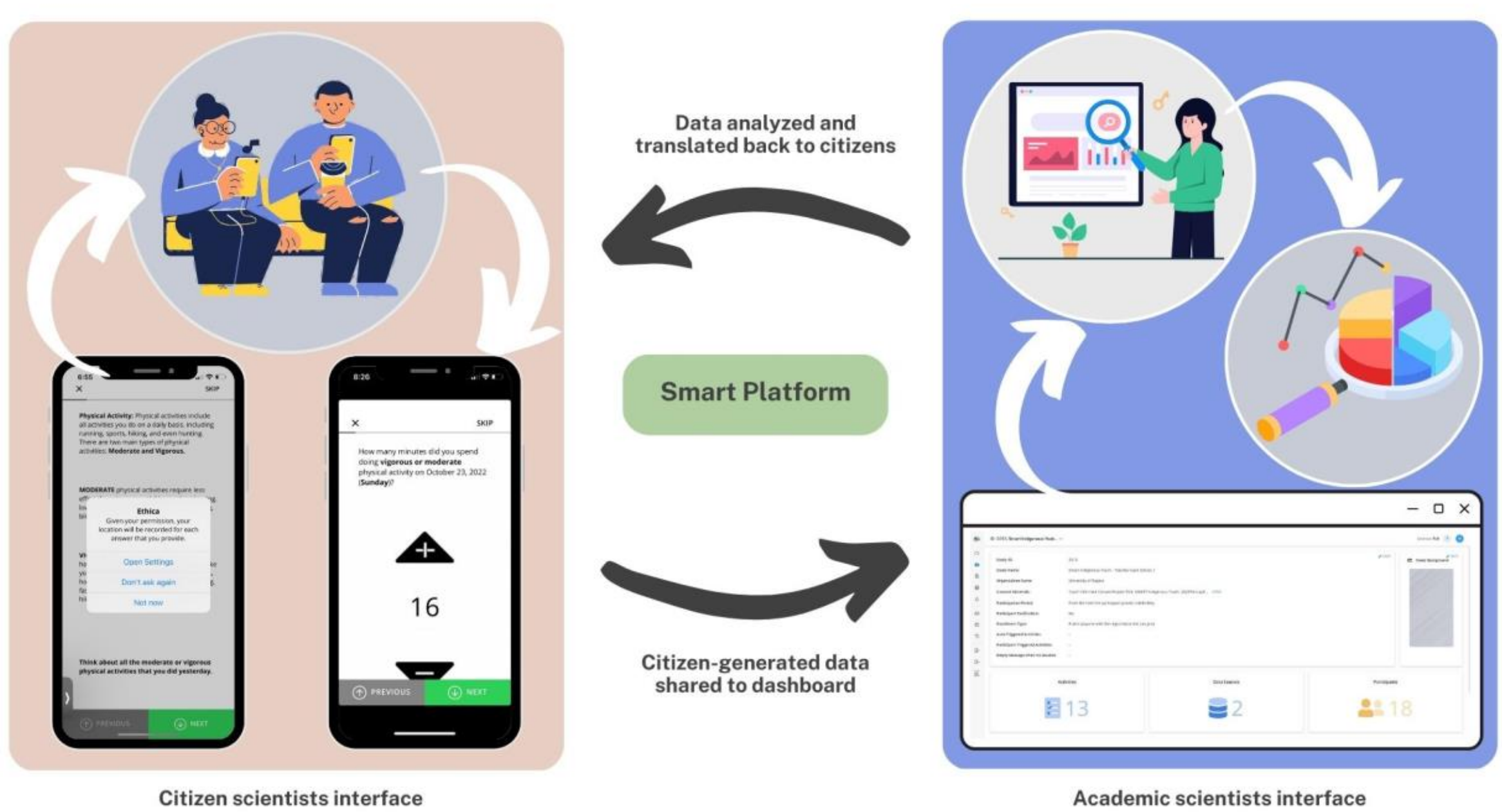

As part of the Smart Platform, a custom-built smartphone application (app) was downloaded by participants using both iOS and Android smartphones. The app enables our research team to engage with participants over eight consecutive days [53] during this study. Through the app, participants reported their MVPA data, behavioural, contextual, demographic and social factors influencing their PA behaviour [24,51,53,54]. The following data were derived from the surveys deployed through the app: peer and family support for PA, friends committed to drug-free sociodemographic characteristics, and strength training of youth citizen scientists. Before any data collection, all participants were required to confirm their age and complete the informed consent form via the app. On day 1 of joining the study, participants were required to provide responses to the modified cross-sectional validated survey measure. Additionally, they were asked to provide responses to mEPAs from day 1 to day 8, including weekdays and weekends through an in-app time-triggered nudge.

**Participants**

Participants in this study are "citizen scientists" because they not only collaborated with the researchers to co-design this study, but also actively engaged with the researchers at all stages of the study process [39]. They played an important role in shaping data collection and driving knowledge translation [39]. For instance, during data collection, youth were able to communicate with the researchers in real-time via the app to provide feedback on data collection strategies. While this feedback enabled adjustment to the study, it also facilitated real-time improvement by the researchers. The optimal sample size calculation adopted for this study was based on the study design [55,56]. By conducting a sample size calculation at a 90% confidence level with a 5% margin of error, it was determined that a sample size of 273 participants was required for the study. However, a total of 808 youth (13 to 21 years old) from five high schools were recruited to participate in this study between August 31st and December 31st, 2018. Youth were recruited through engagement sessions held in various public and catholic high schools in Regina, Saskatchewan, Canada. Using convenience sampling method, all eight high schools in the city of Regina were approached to participate in this study. Five of the eight high schools approached agreed to participate in this study. The participation rate of students in the high schools that agreed to participate was over 88% [57]. Through collaborations with school administrators, the research team scheduled in-person recruitment sessions in schools. During these recruitment sessions, the research team described the study, demonstrated how to use the custom-built app, which included demonstrations of completing mEPAs among other tasks. The research team also answered queries and concerns of youth, and assisted youth in downloading the app onto their respective smartphones. Informed consent was provided by all youth who participated in the study through the app. Implied informed consent was obtained from parents and caregivers of participants who were between 13 and 16 years old prior to the in-person scheduled recruitment sessions. All youth were provided with a one-week free pass to the local YMCA, which is valued at $25 [51].

## Measures

### MVPA (dependent variables)

On the first day of the study, participants were nudged through a smartphone time-triggered nudge within the app to provide retrospective MVPA data (over the previous seven days). The retrospective MVPA data were collected through a survey adapted from three validated self-reported measures: the international physical activity questionnaire, the simple physical activity questionnaire, and the global physical activity questionnaire [3,58,59]. The adaptation in the study allowed participants to report MVPA accumulation over the past seven days preceding their enrollment in the study, regardless of when they joined the study (**S2 Figure 2**) [60]. For example, if a youth became a part of the study on August 31, they were prompted by the app to report their MVPA accumulation for the days of August 30, 29, 28, 27, 26, 25, and 24. Based on the number of days the participants reported engaging in MVPA in the past week, the average frequency of MVPA was calculated.

**S2 Figure 2**: Digitally deployed retrospective MVPA survey.

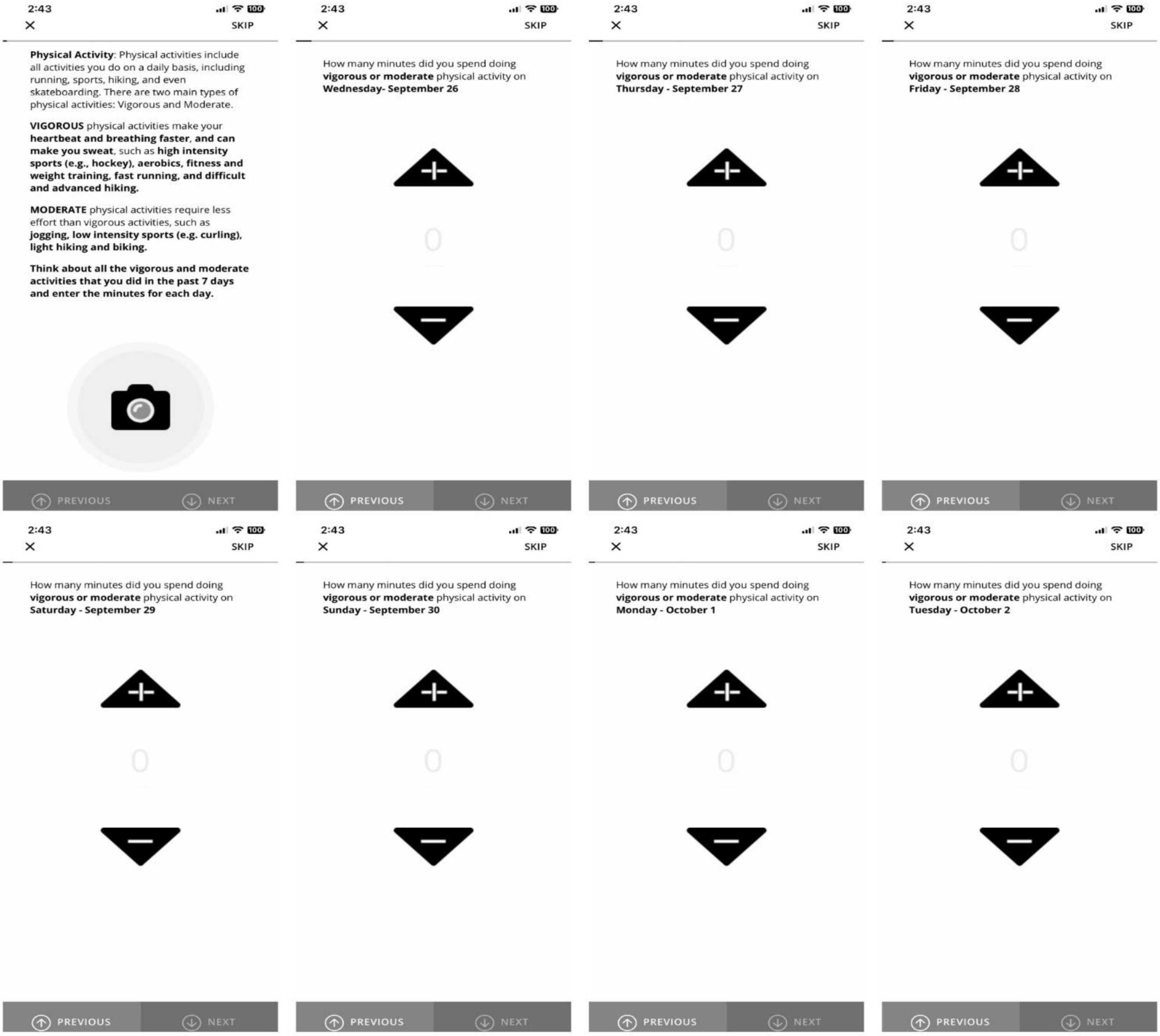

The first screen in **S2 Figure 2** defines what constitutes MVPA, while the following screens present the retrospective survey to provide citizen scientists with the opportunity to report their MVPA accumulation over the previous seven days, beginning on the first day of joining the study. Based on these responses, the frequency of MVPA for that day (will be referred to as retrospective MVPA frequency) was derived. The frequency of MVPA was defined as any day in which youth reported more than 10 minutes of MVPA following the international physical activity questionnaire data processing guidelines [61].

Prospective MVPA was reported through daily time-triggered mEPAs from day 1 through day 8 of the study period including weekdays and weekends. To ensure adequate time is allowed to report MVPA, mEPAs were automatically deployed to citizen scientist smartphones in the evening between 8:00 PM and 11:30 PM and were set to expire at 1 AM daily [24]. mEPAs had skip patterns to ensure flexibility by enabling citizen scientists to move from one question to another (**S3 Figure 3**) [60].

**S3 Figure 3**: Digitally deployed mEPAs

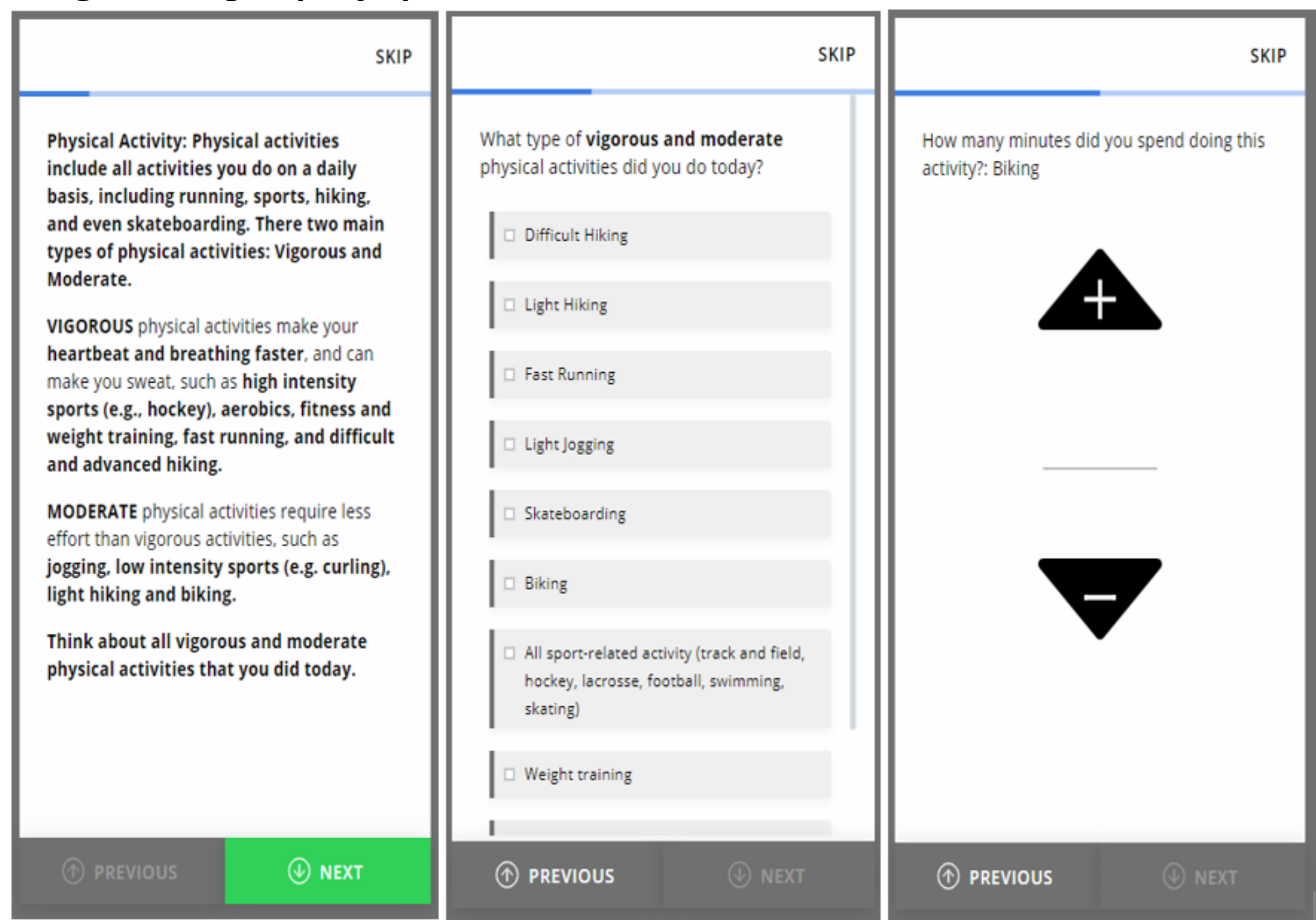

The first screen in **S3 Figure 3** defines what constitutes MVPA, followed by a series of mEPA questions including: "What type of physical activities did you do today?" (Multiple choice) and "How many minutes did you spend doing this activity?" (Open-ended). From these questions, the frequency of MVPA for that day was derived by selecting a cut point to include engagement in at least one MVPA for 10 or more minutes (will be referred to as mEPA MVPA frequency). The dependent variables for this study included both retrospective and mEPA MVPA frequency.

**Peer support for MVPA (independent variables)**

Youth were asked to consider their closest friends in the last 12 months when answering the question regarding peer support of MVPA. Peer support for MVPA was captured with the question: "How many of your closest friends are physically active?" with the six response options: "none of my friends", "1", "2", "3", "4", or "5 of my friends". Responses for peer support for MVPA were dichotomized into "zero physically active friends" corresponding to "none of my friends" and "at least 1 active friend" corresponding to "1", "2", "3", "4" or "5 of my friends".

**Family support for MVPA (independent variables)**

Family support for MVPA was captured by asking Youth "How many times in a week an adult in your household encourages PA" with response options: "Never", "1-2 times per week", "3-4 times per week", "5-6 times per week" and "Daily". Responses were categorized into "Daily", "5-6 times per week", and "4 times or less per week" corresponding to "Never", "1-2 times per week", and "3-4 times per week".

**Friends drug free (independent variables)**

Youth were asked to consider their friends when answering this question. Youth were asked "How many of your closest friend are committed to drug-free" with five response categories: "None of my friends", "1 of my friends", "2 of my friends", "3 of my friends" and "4 of my friends". These responses were categorized into "None of my friends", "1 of my friends", "2 of my friends" and "3 or more of my friends" which corresponds to "3 of my friends" and "4 of my friends".

**Distance from home to school (independent variables)**

Youth were asked: "To the best of your knowledge, how far is your school from home" with response categories: "Less than 2km", "Less than 5km", "Less than 10km", "Less than 20 km", and "More than 20 km". The responses were dichotomized into "Less than 5 km" which corresponds to "Less than 2km", and "Less than 5km", and "5 km or more" which corresponds to "Less than 10km", "Less than 20 km" and "More than 20 km".

**Sociodemographic covariates**

Gender was captured by asking participants "What is your gender?", with five response options: "male", "female", "transgender", "other (please specify)", and "prefer not to disclose". In order to address the low counts within the categories, responses that were "transgender," "other," and "prefer not to disclose" were combined into a single category. Information on parental education was collected by asking them about the highest level of education of one of their parents or guardians through six response options: "elementary school", "some secondary/high school", "completed high school", "some post-secondary (university/college)", "received university or college degree /diploma", and "does not apply". Each of the six responses was organized into four categories of parental education: 1) "elementary school" corresponds to "elementary school or below", 2) "some secondary/high school" and "completed high school" corresponds to "at least secondary school" 3) "some post-secondary (university /college)", "received university or college degree/diploma" corresponds to "university and above" and "does not apply".

The ethnicity of participants was also captured through the following response options: "First Nations", "Dene", "Cree", "Metis", "Inuit", "African", "Asian", "Canadian", "Caribbean/West Indian", "Eastern European", "European", "South Asian", "other", and "Mixed". The responses were grouped into four categories: 1) "Indigenous" which corresponds to "First

Nations", "Dene", "Cree", "Metis", "Inuit", 2) "Canadian", 3) "mixed", and 4) "visible minorities". The visible minorities include "African", "Asian", "Caribbean/West Indian", "Eastern European", "European", "South Asian", and "other" categories. The category "visible minorities" was established as a result of the low number of responses within individual ethnic categories. Youth were also asked if they had a part-time job, with the following response options: "Yes" or "No".

**Strength training (independent variables)**

Engagement in strength training was measured by asking participants "On how many days in the last 7 days did you do exercises to strengthen or tone your muscles? (e.g., push-ups, sit-ups, or weight-training)". Youth were provided with eight response options including: "0 ", "1 ", "2 ", "3 ", "4 ", "5 ", "6 ", or "7 days''. Each of the responses was dichotomized into "0 days of strength training" corresponding to 0 days and "at least 1 day of strength training" corresponding to "1", "2", "3", "4", "5", "6", or "7 days".

**Data and risk management**

The app used throughout the study was built to ensure confidentiality, data safety, and security. All data were encrypted before being streamed to a secured cloud server. The custom-built smartphone app had permissions which restrict access to personally identifiable information present in youth smartphones (e.g., site visited, location, and contact list). Youth data was encrypted using a hash algorithm. During in-person recruitment sessions, the risk and privacy management options were made clear to participants through the informed consent form. Within the app, youth were given the option to drop out or pause data collection anytime during the study period. Also, participants could decide to upload their data only when they had access to WI-FI or when their smartphones were plugged into a power source. Clear instructions were provided within the app for participants who wished to withdraw from the study at any point in time [39,51].

## Data Analyses

The inclusion criteria for data to be considered for analyses was that participants had to provide MVPA data using both retrospective and prospective measures to ensure methodological rigour. Data analyses for this study were conducted using R 4.2.1, an open-source statistical tool. Frequencies and percentages were used to describe the categorical independent variables and chi-square tests were used to test the significance difference in a proportion of independent variables included in this study. Wilcoxon sign rank test was used to ascertain the difference between retrospective and mEPA MVPA frequency. Further, Poisson regression analyses were used to assess factors associated with retrospective and mEPA MVPA frequency. Covariates including peer and family support [62,63], strength training [64,65], distance and friends' drug-free were included in this study based in previous related literatures. Cases with missing data points were omitted from data analyses [66]. All results were considered statistically significant at $p < 0.05$.

## Results

This study involved the participation of 808 youth (13-21 years); however, when the cross-sectional survey was deployed at baseline, 436 participants provided complete data using the retrospective measure for capturing MVPA. Moreover, after excluding participants who did not provide complete MVPA data using the prospective measure (mEPA), the final sample size was 68 participants. The summary statistics of the participants (N=68) are summarized in **Table 1**. Youth were predominantly males (40.30%) of youth, with 56.72% being females, and 2.98% of them reported being transgender, other, or preferring not to reveal. The predominant ethnicity was Canadian (37.31%), while 35.82% identified as mixed, 25.37% identified as visible minority, and 1.50% as Indigenous. In terms of socioeconomic status, most youth (72.06%) reported that one of their parents had at least a university degree. For strength training, 81.54% of participants reported that they engaged in strength training at least once, while 18.46% reported they had never engaged in strength training. Additionally, a majority of youth reported having at least one or more physically active friends (88.88%). Further, most participants (70.15%) reported not having a part-time job.

**Table 1**: Summary statistics of youth participating in this study

| **Dependent Variables** | **Average Frequency per week** | | | | | |
|---|---|---|---|---|---|---|
| Retrospective MVPA | 5.06 (2.47) | | | | | |
| mEPA MVPA | 2.10 (1.78) | | | | | |
| | | | | | | |
| **Independent Variable** | **Mean (S.D)** | | | | | |
| Age | 15.45 (1.46) | | | | | |
| | | | | | | |
| **Independent Variables** | **Baseline (N=436)** | | | **Final Sample (N=68)** | | |
| **Gender** | **Frequency** | **Percentage** | **Sig** | **Frequency** | **Percentage** | **Sig** |
| Male | 161 | 38.5 | | 27 | 40.30 | p<0.000 |
| Female | 233 | 55.8 | p<0.000 | 38 | 56.72 | |
| Transgender / Other / Prefer not to disclose | 24 | 5.7 | | 2 | 2.98 | |
| Total | 418 [a] | 100 | | 67 [a] | 100 | |
| | | | | | | |
| **Ethnicity** | | | | | | |
| Indigenous | 21 | 5.0 | | 1 | 1.50 | p<0.000 |
| Canadian | 166 | 39.8 | p<0.000 | 25 | 37.31 | |
| Mixed | 124 | 29.7 | | 24 | 35.82 | |
| Visible minority | 106 | 25.5 | | 17 | 25.37 | |
| Total | 417 [a] | 100 | | 67 [a] | 100 | |
| | | | | | | |
| **School** | | | | | | |
| 1 | 110 | 25.3 | | 16 | 23.5 | p<0.000 |
| 2 | 74 | 17.1 | | 4 | 5.90 | |
| 3 | 50 | 11.5 | p<0.000 | 4 | 5.90 | |
| 4 | 78 | 18.0 | | 2 | 32.35 | |
| 5 | 122 | 28.1 | | 22 | 32.35 | |
| Total | 434 [a] | 100 | | 68 | 100 | |
| | | | | | | |

| **Parental education** | | | | | | |
|---|---|---|---|---|---|---|
| At least secondary school | 91 | 21.0 | p<0.000 | 14 | 20.59 | |
| University and above | 282 | 65.1 | | 49 | 72.06 | |
| Does not apply | 48 | 11.1 | | 5 | 7.35 | |
| Total | 433 [a] | 100 | | 68 | 100 | |
| | | | | | | |
| **Distance from home to school** | | | | | | |
| Less than 5 km | 248 | 57.54 | p = 0.001 | 42 | 62.69 | p =0.038 |
| 5 km or more | 183 | 42.46 | | 25 | 37.31 | |
| Total | 431 [a] | 100 | | 67 [a] | 100 | |
| | | | | | | |
| **Strength training** | | | | | | |
| Zero days of strength training | 97 | 23.5 | p<0.000 | 12 | 18.46 | p<0.000 |
| At least one day of strength training | 315 | 76.5 | | 53 | 81.54 | |
| Total | 412 | 100 | | 65 [a] | 100 | |
| | | | | | | |
| **Peer support for MVPA** | | | | | | |
| Zero active friends | 47 | 11.3 | p<0.000 | 8 | 12.12 | p<0.000 |
| At least one active friend | 368 | 88.7 | | 58 | 88.88 | |
| Total | 415 [a] | 100 | | 66 [a] | 100 | |
| | | | | | | |
| **Family support for MVPA** | | | | | | |
| 4 times or less per week | 333 | 78.35 | P<0.000 | 52 | 78.79 | p<0.000 |
| 5-6 times per week | 10 | 2.35 | | 2 | 3.03 | |

| Daily | 82 | 19.30 | | 12 | 18.18 | |
|---|---|---|---|---|---|---|
| Total | 425 [a] | 100 | | 66 [a] | 100 | |
| | | | | | | |
| **Friend drug free** | | | | | | |
| None of my friends | 113 | 26.97 | p<0.000 | 14 | 20.90 | p<0.000 |
| 1 of my friends | 49 | 11.69 | | 8 | 11.94 | |
| 2 of my friends | 53 | 12.65 | | 8 | 11.94 | |
| 3 or more of my friends | 204 | 48.69 | | 37 | 55.22 | |
| Total | 419 [a] | 100 | | 67 | 100 | |
| | | | | | | |
| **Part time job** | | | | | | |
| Yes | 147 | 37.8 | p<0.000 | 20 | 29.85 | p<0.000 |
| No | 242 | 62.2 | | 47 | 70.15 | |
| Total | 389 [a] | 100 | | 67 | 100 | |

[a] Some youth did not provide a response to this question.

The summary statistics and the Wilcoxon sign rank test of the outcome variables for this study are presented in **Table 2**. Youth reported engaging in MVPA for an average of 5.06 times per week retrospectively and 2.10 times per week prospectively. The Wilcoxon signed rank test (p < 0.001) indicated a statistically significant difference between the mean frequency of MVPA reported retrospectively and prospectively via mEPAs.

**Table 2**: Wilcoxon signed rank test showing the difference between retrospective MVPA and mEPA MVPA frequency.

| | Average (Frequency per week) | Minimum (Frequency per week) | Maximum (Frequency per week) | N | Wilcoxon signed rank test. (p-value) |
|---|---|---|---|---|---|
| **MVPA frequency** (Retrospective MVPA frequency) | 5.06 | 0 | 7 | 68 | p < 0.000 |
| **mEPA MVPA frequency** (Prospective MVPA frequency) | 2.10 | 0 | 7 | 68 | |

The Poisson regression analyses showing the association between retrospective MVPA frequency (model 1) and prospective MVPA frequency (model 2) with sociodemographic and contextual factors are presented in **Table 3**. In the retrospective MVPA model, two contextual variables were found to be significantly associated with retrospective MVPA frequency. For instance, youth who attended school 3 (β = -1.347, [C.I.] = -2.383, -0.311, p–value = 0.010) reported a lower frequency of MVPA in comparison to youth who attended school 1. Moreover, youth who engaged in at least one day of strength training (β = 0.477, 95% [C.I.] = 0.042, 0.912, p–value = 0.031) reported a higher frequency of MVPA in comparison to youth who did not engage in strength training.

Contrastingly, in the mEPA model (i.e., prospective MVPA: model 1), several contextual variables were found to be significantly associated with the frequency of MVPA. With respect to family support for MVPA, youth with a family member who encouraged them to participate in PA 5 to 6 times per week (β = 1.294, 95% [C.I.] = 0.282, 2.306, p–value = 0.012) reported a higher frequency of MVPA than youth whose family members encouraged MVPA 4 times or less per week. In contrast, youth who had family member who encouraged them to participate in MVPA daily (β = -0.699, 95% [C.I.] = -1.320, 0.078, p–value = 0.027) reported a lower frequency of MVPA in comparison to youth whose family members encouraged MVPA 4 times or less per week. Further, youth who had two (β = 0.870, 95% [C.I.] = 0.094, 1.645, p–value = 0.028) or three (β = 0.924, 95% [C.I.] = 0.322, 1.526, p–value = 0.003) friends who were drug-free reported higher MVPA in comparison to youth who had no friend who were drug-free. In terms of peer support, youth who had at least one active friend (β = -0.898, 95% [C.I.] = -1.633, -0.162, p–value = 0.017) reported a lower MVPA frequency in comparison to youth who had no physically active friends. Additionally, youth who did not have a part time job (β = -0.615, 95% [C.I.] = -1.173, -0.057, p–value = 0.031) reported a lower frequency of MVPA in comparison to youth who had a part time job. Finally, youth whose distance from school to their home was 5 km or more (β = -0.615, 95% [C.I.] = -1.173, -0.057, p–value = 0.031) reported a higher frequency of MVPA in comparison to youth whose school is less than 5 km from their homes.

**Table 3:** Poisson regression models showing the association between retrospective and mEPA MVPA frequency and sociodemographic and contextual factors [a].

| Variable | **Model 1: Retrospective MVPA frequency** | **Model 2: mEPA MVPA Frequency** |
|---|---|---|
| **School** | | |
| **1 (Ref)** | | |
| 2 | 0.317 (-0.229, 0.863) | 0.694 (-0.077, 1.465) |
| 3 | **-1.347** (-2.383, -0.311)** | -0.026 (-0.999, 0.948) |
| 4 | 0.375 (-0.112, 0.863) | -0.542 (-1.306, 0.223) |
| 5 | 0.214 (-0.253, 0.682) | -0.048 (-0.800, 0.704) |
| **Family Support for MVPA** | | |
| **4 times per week (Ref)** | | |
| 5-6 times per week | 0.586 (-0.132, 1.305) | **1.294** (0.282, 2.306)** |
| Daily | 0.178 (-0.158, 0.513) | **-0.699** (-1.320, -0.078)** |
| **Friends Drug free** | | |
| **None of my friends (Ref)** | | |
| 1 of my friends | 0.378 (-0.151, 0.907) | 0.666 (-0.116, 1.448) |
| 2 of my friends | 0.231 (-0.273, 0.735) | **0.870** (0.094, 1.645)** |
| 3 or more of my friends | 0.191 (-0.178, 0.560) | **0.924** (0.322, 1.526)** |
| **Strength Training** | | |
| **Zero days of strength training (Ref)** | | |
| At least one day of strength training | **0.477** (0.042, 0.912)** | 0.108 (-0.510, 0.726) |
| **Peer support for MVPA** | | |
| **Zero physically active friends (Ref.)** | | |

| At least one active friend | 0.402 (-0.090, 0.893) | -0.898** (-1.633, -0.162) |
|---|---|---|
| **Part time Job** | | |
| **Yes (Ref.)** | | |
| No | 0.061 (-0.295, 0.418) | -0.615** (-1.173, -0.057) |
| **Distance** | | |
| **Less than 5 km (Ref)** | | |
| 5 km or more | -0.053 (-0.349, 0.243) | 0.486** (0.001, 0.971) |
| Constant | -2.133 (-5.016, 0.749) | 1.060 (-3.563, 5.683) |
| Observations | **62**[b] | **62**[b] |

*** $p < 0.05$

[a] *Both models controlled for Age, Gender, Parental education, and Ethnicity.*

[b] *Six observations were deleted due to missing data*

**Discussion**

This study was carried out to ascertain if there is a significant difference between the frequency of MVPA assessed through a digitally deployed retrospective survey, modified from three validated PA questionnaires, and prospectively through mEPAs. Understanding the frequency of MVPA is essential for NCD prevention and management as it has not only been associated with improved health outcomes in the youth population [67], but it is also a key component of existing PA guidelines for children and youth [68]. Further, this study explored various factors associated with the frequency of MVPA reported retrospectively and through mEPAs in the same cohort of participants (13 to 21 years old).

The main finding of the study was that there was a significant difference between the frequency of MVPA reported retrospectively and the frequency of MVPA reported via mEPAs. In particular, youth reported a higher frequency of MVPA retrospectively than via mEPAs. This difference could be due to recall biases inherent in retrospectively reported PA as pointed out by previous research [69,70]. Recall bias has been reported to be a major contributor to the overreporting of MVPA [71], especially when the instrument is used over a longer recall period. However, not all self-reported retrospective PA instruments are prone to recall biases. For instance, some studies have pointed out that time-use diaries have the potential to reduce recall bias in PA assessments [72,73]. Additionally, self-reporting bias [74], has been reported to be a major shortcoming of retrospective self-assessed behaviour [75]. For instance, a study reported that students in grade 9 to 12 tend to over-report their height but under report their age [76]. Further, external bias related to social desirability has also been found to modify the reporting of self-assessed behaviour [74]. Social desirability bias is the propensity to self-report behaviour in a way that is viewed as desirable by society [17]. Although social desirability bias has been shown to be a barrier to retrospective in-person data collection [77], digitally deployed surveys can reduce this bias, owing to its anonymous nature [78].

Prospective assessments, characterized by a shorter duration of reporting periods, can mitigate recall biases inherent in retrospective surveys [49]. In particular, collecting PA data close to the time it occurred reduces the reliance on memory recall over extended periods, thus, resulting in more reliable and accurate data [79–81]. Moreover, prospective assessments of PA, enable more frequent data collection, which provides more granular data in near-real-time [82]. Near real-time data collection minimizes the potential distortion in data reporting that may arise from retrospective recall bias [83]. Overall, the short reporting periods of prospective assessments not only offer a means of minimizing recall biases but also facilitates more accurate, timely, and comprehensive data collection, thereby enhancing the reliability and validity of research findings [80]. These findings on the reported frequency of MVPA are a valuable addition to the existing literature as previous studies have not examined the differences between prospective and retrospective PA reporting among youth, particularly the frequency of MVPA.

With the widespread availability of digital devices and the increasing digital literacy among youth, digital citizen science approaches for surveillance of MVPA has immense potential to manage and prevent NCDs [24,39,52,84–86]. In carrying out this study, youth citizen scientists were engaged using their own smartphones, an approach that prioritizes ethical population health surveillance using digital citizen science approaches [39]. However, in taking this digital citizen science approach to understanding MVPA accumulation via mEPAs, it was important to investigate the association of MVPA frequency with sociodemographic, contextual, and behavioural factors among youth [60]. Several sociodemographic and contextual factors were considered to further understand the difference in the frequency of MVPA reported retrospectively (model 1) and via mEPAs (model 2). Our study found five factors to be significantly associated with MVPA frequency in the mEPA model, while there were only two factors significantly associated with MVPA frequency in the retrospective model. The observed difference in the number of significant associations between the mEPA and retrospective models suggests that mEPA may be able to more effectively identify the factors associated with MVPA among youth compared to retrospective measures.

In the retrospective model, school enrolment was found to be significantly associated with the frequency of MVPA. However, this association was not found to be significant in prospectively model. In particular, our study found that youth who attended School 3 reported a lower frequency of MVPA in comparison to youth who attended School 1. This difference could be attributable to several factors, including differences in school sports programs [87,88], physical education curricula [89], and access to PA facilities within educational institutions[90,91]. In particular, previous studies have reported that school settings such as facilities, space, equipment, and leadership support could either serve as barriers or facilitators of student healthy behaviour among students [54,92]. These study findings on the association of potential school environment with MVPA frequency reported via retrospective measures should be further explored considering the importance of school environment on youth MVPA behaviour.

Our study also found that youth who engaged in strength training reported a higher frequency of MVPA using mEPAs. Previous studies have found that individual effort in strength training was associated with retrospectively reported PA duration among youth [93,94]. However, there is little evidence linking MVPA frequency reported via mEPAs with strength training among youth. As strength training relies on both the frequency and duration of activity [95], prospective measures (mEPAs) are more appropriate for its measurement as they are known to reduce recall bias [96,97] and ensure activities are recorded in near real-time, and real-world settings [98,99].

In contrast with the retrospective model, our study found that various family and peer support factors were associated with frequency of MVPA in the mEPA model. For instance, youth who had an adult in the household who encouraged PA regularly (5 to 6 times per week) were associated with a higher frequency of MVPA, in comparison to youth who had an adult that encouraged PA 4 times or less per week. These findings align with previous studies which found that family and parent encouragement of PA was associated with more minutes of PA among youth [100–102]; however, their study focused on PA duration, and not frequency of MVPA. Consistent encouragement from adults may provide youth with the motivation and support needed to overcome barriers and maintain a regular exercise regimen. On the contrary, our study interestingly found that youth reporting that an adult encouraged PA daily was associated with less frequency of MVPA. This association may be attributable to youth perceiving daily PA encouragement as overly pressuring or controlling, leading to resistance of MVPA [103]. In addition to family encouragement, our study found that having one or more physically active friends was associated with lower frequency of MVPA, which contrasts with previous studies conducted among youth [27]. One possible explanation for this could be that having several physically active friends may make some youth feel inadequate through social comparison [104], making them feel inferior or less motivated to engage in MVPA themselves. However, future studies examining the causal pathways between social support factors, including physically active friends, and frequency of MVPA are imperative for better understanding the barriers and facilitators of PA among youth.

The mEPA model also found that distance to school and having a part time job were significantly associated with the frequency of MVPA. In particular, attending a school that was greater than five kilometres away from home was associated with a higher frequency of MVPA compared to those whose schools were within five kilometers from home. This difference could be attributable to youth who live further from school engaging in more active school transportation [36]. Previous evidence has found short distances to school to be associated with a higher likelihood of active transport. However, the higher probability of active transport was offset by the lower PA associated with shorter distance commutes [105]. Additionally, our study found that youth who had no part time job had a lower frequency of MVPA compared to those with a part time job. This difference could potentially be explained by work-related PA, as youth who have a part time job may engage in PA as a part of their job. Moreover, these results align with previous evidence which indicates that youth with a part-time job engage in more MVPA than youth who do not have a part time job [106].

There is considerable evidence that mEPA is a feasible and reliable methodology for studying daily behaviours in youths' natural environments, and in near real-time [97,107,108]. In addition, mEPAs advance novel measurement strategies that go beyond traditional retrospective self-reports. For instance, the repeated measurement of MVPA made possible by mEPAs facilitates a robust assessment of associations between MVPA and its correlates [97]. Furthermore, mEPA methodologies have been reported to enhance novel PA assessment strategies that can be rapidly implemented in real-time. Thus, these approaches can facilitate improved data quality and compliance among study participants [108]. However, there is little evidence on the surveillance of MVPA frequency among youth using mEPAs. Our study addresses critical gap by examining the differences in MVPA reporting via retrospective and mEPAs within the same cohort of individuals.

**Strengths and Limitations**

This study adds novel evidence to our understanding of MVPA frequency among youth using advanced digital technologies. Due to the adoption of ubiquitous devices (smartphones) in the deployment of PA surveys, this study depicts ethical digital citizen science approaches, where citizen-owned devices can be used to monitor MVPA behaviour both prospectively and retrospectively (mEPAs). Limitations of this study include combining a cross-sectional design with mEPA measures, leading to the loss of important data points, and limiting the ability to establish causal relationships between the variables examined. Nevertheless, a key advantage of using digital epidemiological approach is the ability to recruit large number of individuals and engage them in real-time. However, the burden on the participants needs to be balanced.

For instance, in this study a significant proportion of youth (85.75%) did not provide complete data for prospective measures, which reduced our sample size. Future studies should minimize burden on participants by limiting data collection to only MVPA to improve compliance. In addition, participants were engaged prospectively using mEPAs across only eight days to enhance comparability with retrospective surveys. Due to logistic and compliance purposes, our study obtained retrospective MVPA data covering seven days preceding the prospective data collection phase. Further, since the modified survey was derived from three validated surveys, the validity and reliability were implied from the validated surveys. Additionally, this study assumes that both measures of retrospective and prospective PA assume typical MVPA habit. Future research endeavour should be carried out in multiple jurisdictions, while adopting longitudinal designs and integrating mEPAs with objective (accelerometer and pedometer) data collection to better understand MVPA frequency among youth. Moreover, although we were able to recruit more participants than required by our sample size calculation, the number of participants included in the analyses was lower due to participants not providing complete data throughout the study period. Future studies should enhance sample size for the purpose of improving the generalizability of study findings.

## Conclusion

The main finding of this study showed that there was a significant difference between the frequency of MVPA reported retrospectively versus prospectively. Additionally, ethnicity, parental education, and strength training were found to have significant associations with prospectively reported MVPA while no significant association were found with the retrospectively reported MVPA. Digital citizen science approaches can be used to ethically monitor MVPA frequency, not only retrospectively, but also prospectively using mEPAs. This study highlights the important differences in using retrospective vs. mEPAs measures, findings which are useful to develop appropriate prevention and management strategies for NCDs.

Overall, this study not only transforms how we utilize digital devices ethically to understand human behaviour but also provides insights to develop potential behavioural interventions that can be deployed using the same digital devices.

**Acknowledgments**

The authors acknowledge the entire Digital Epidemiology and Population Health Laboratory team (DEPtH) for their unwavering support, as well as the Canadian Institute of Health Research for their support to the DEPtH Lab and the Smart Platform.